\documentclass[namedreferences]{solarphysics}
  \usepackage[hyperref,optionalrh,solaromanenum]{spr-sola-addons} 
  \usepackage{graphicx}                    
  \usepackage{color}                       

  \chardef\us=`\_

\begin{document}

\begin{article}
\begin{opening}

\title{Meridional Circulations of the Solar Magnetic Fields of Different Strength}


 \author[addressref=,
 email={bilenko@sai.msu.ru}]
 {\inits{I. A}\fnm{Irina A.}~\lnm{Bilenko}\orcid{0000-0002-9543-0542}}


 %
 \runningauthor{Irina A. Bilenko}
 \runningtitle{Meridional Circulation of Magnetic Fields}

\address{Sternberg Astronomical
 Institute Moscow M.V. Lomonosov State University,
 Universitetsky pr.13, Moscow, 119234, Russia}

\begin{abstract}
The meridional circulation of the solar magnetic fields
in Solar Cycles 21\,--\,24 was considered.
Data from both ground-based and space observatories were used.
Three types of time-latitude distributions of photospheric
magnetic fields and their meridional circulations were
identified depending on the magnetic field intensity.
(i) low-strength magnetic fields. Positive- and negative-polarity
magnetic fields were distributed evenly across latitude and
they weakly depended on the magnetic fields of active regions and
their cycle variation;
(ii) medium-strength magnetic fields.
For these positive- and negative-polarity magnetic fields a wave-like,
pole-to-pole, antiphase meridional circulation
with a period of $\approx$22 years was revealed.
The velocities of meridional flows  were slower
at the minima of solar activity, when they were at high  latitudes
in the opposite hemispheres,
and maximal at the solar maxima,
when the positive- and negative-polarity waves crossed the equator.
The meridional circulation of these fields reflects the solar global
magnetic field dynamics and determines the solar polar field reversal;
(iii) high-strength (local, active region) magnetic fields.
They were distributed symmetrically in the Northern and Southern hemispheres.
The magnetic fields of active regions were formed only
during the periods when the positive- and negative-polarity waves
of medium-strength magnetic fields approached at low latitudes.
Magnetic fields of both leading and following sunspot
polarity migrated from high to low latitudes.
The meridional-flow velocities of high-strength magnetic fields
were higher at the rising and maxima phases than at the minima.
Some of the  high-latitude active region magnetic fields
were captured by the second type meridional circulation flows
and transported along with them to the appropriate pole.
But the magnetic fields of active regions are not
the main ones in the solar polar field reversal.
The results indicate that high-strength magnetic fields
were not the main source of weak ones.
The butterfly diagram is the result of a superposition
of these three types of  magnetic field time-latitude
distributions and their meridional circulation.
The results suggest that different strength magnetic fields
have different sources of their generation and cycle evolution.
\end{abstract}

%

\end{opening}


\section{Introduction}
    \label{S-Introduction}

Solar plasma meridional circulation play a major role
in the solar magnetic field dynamics.
Meridional flows globally redistribute the solar plasma, heat,
angular momentum, and magnetic fields in all layers from convective zone
to the solar corona.
Meridional flows are important in explaining the solar dynamo and
differential rotation.
They play a key role in the flux-transport dynamo models
(\citealp{Sheeley2005, Jiang2014, Hanasoge2022}).
Cycle variations of the meridional flow have been suggested to explain
the changes in the amplitudes and lengths of the solar activity cycles
\citep{Charbonneau2020, Karak2023}
and they also used for their prediction \citep{Petrovay2020}.
Solar meridional circulation is considered to be an axisymmetric
flow system, extending from the equator to the poles
with $\approx$20~m~s$^{-1}$ at the solar surface.

Various methods and traces are used to identify and measure
meridional flows.
Poleward flow has been detected using the Doppler method.
\cite{Duvall1979} found a poleward flow of 20~m~s$^{-1}$ approximately
constant over the latitude range of 10$^{\circ}$\,--\,50$^{\circ}$.
Studying giant velocity features  about 15$^{\circ}$ in latitude and 30\,--\,60$^{\circ}$
in longitude with amplitudes around 40~m~s$^{-1}$ on the solar
surface, \cite{Howard1979} have also found a meridional flow
toward the poles of the order of 20~m~s$^{-1}$.
Using doppler velocities \cite{Hathaway1996} determined an antisymmetric
flow towards the poles of 20~m~s$^{-1}$ about the equator in both hemispheres.
For bright Ca+-mottles \cite{Schroeter1975} have found a systematic meridional
motion of about 0.1~km~s$^{-1}$ for latitudes around 10$^{\circ}$.
\cite{Topka1982} shown that polar filaments track
the boundary latitudes of the unipolar
magnetic regions and drift poleward with the regions at about 10~m~s$^{-1}$.
They noted that such a filament motion cannot be explained by diffusion alone,
and that a poleward meridional flow carries magnetic flux
of both polarities along with it.

Small magnetic elements are often used to determine
the velocity of meridional flows \citep{Rightmire2012}.
\cite{Komm1993} found poleward meridional flow velocity of the order
of 10~m~s$^{-1}$ in each hemisphere for small magnetic elements
using high-resolution magnetograms taken from 1978 to 1990
with the NSO Vacuum Telescope on Kitt Peak.
The meridional flow was found to change during solar cycle.
They showed that meridional flow increased in amplitude from the equator,
reached a maximum of 13.2~m~s$^{-1}$ at 39$^{\circ}$, and decreased poleward.
They found no significant hemispheric asymmetry and
no equator ward migration.
However, analyzing the Mt. Wilson magnetograms, \cite{Snodgrass1996}
found that for latitudes near the equator the flow was towards the equator.

Meridional flows were found to change during solar cycles.
\cite{Hathaway2010} using SOHO data showed that
the average flow speed of magnetic features was poleward
at all latitudes up to 75$^{\circ}$.
They found that the flow was faster at sunspot cycle minimum than
at maximum and it was changed from cycle to cycle.
In \cite{Hathaway2014} it was also found that the meridional flow
speed of magnetic elements was fast
at solar cycle minimum and slow at maximum.
The meridional flow weakening on the poleward sides of the
active region latitudes and it was slower at Solar Cycle 23.
They proposed that an inflow towards the sunspot zones was
superimposed on a more general poleward meridional flow.
\cite{Imada2020} found that the average meridional flow in
Solar Cycle 24 was faster than that in Solar Cycle 23.

Meridional motions have also been determined using coronal
bright points from SOHO/EIT, Yohkoh/SXT, Hinode, and SDO/AIA data
\citep{Vrsnak2003, Sudar2016}.
\cite{Vrsnak2003} noted that a velocity pattern of bright point motions
reflected the large-scale plasma flows.
In a complex pattern of meridional motion, they found that
the equator ward flows were dominated at low ($B<10^{\circ}$)
and high ($B>40^{\circ}$) latitudes, and a poleward flow
at mid-latitudes ($B \approx10^{\circ}$\,--\,40$^{\circ}$).
Faster flow tracers had equator ward motion and the slower ones
showed poleward motion.

\cite{Howard1986} studying the meridional motions of sunspots
and sunspot groups found a midlatitude northward flow
with a few hundredths of a degree per day in each hemisphere.
For sunspot groups, a generally poleward motions at higher latitudes
was determined.
\cite{Sudar2014} considered the location measurements
of sunspot  groups covering 1878\,--\,2011.
They found that the meridional motion of sunspot groups is towards
the centre of activity from all latitudes and in all solar cycle phases.
The range of meridional velocities was found to be $\pm1$0 m~s$^{-1}$.

Helioseismology makes it possible to study meridional flows
at different layers in the solar interior \citep{Gizon2004}.
\cite{Chou2001} studying the subsurface meridional flow as a function
of latitude and depth from 1994 to 2000, have found
that the velocity of meridional flow increased
when solar activity decreased.
A new meridional flow component at about 20$^{\circ}$
appeared in each hemisphere as solar activity increased.
At low latitudes, the new flow changed from poleward at solar minimum
to equatorward at solar maximum.
The velocity of the new component increases with depth.
In \cite{Basu2003}, it was revealed  that the meridional flows
showed solar activity-related changes.
The antisymmetric component of the meridional flow
decreased in speed with activity.
\cite{Basu2010} have shown that meridional-flow speed
increases with depth.
For Solar Cycle 23, they found that solar meridional flows
in the outer 2\% of the solar radius
was connected with a flow pattern drifting equator ward in parallel
with the activity belts.
The different flow components was found to have different time dependencies,
and  the dependence was different at different depths.
\cite{Zhao2013} using the helioseismology observations
from the SDO/HMI, found the poleward meridional flow of
a speed of 15~m~s$^{-1}$ from the photosphere to about 0.91~R${_\odot}$
and an equator ward flow of a speed of 10~m~s$^{-1}$
between 0.82 and 0.91~R${_\odot}$.
They also found that the meridional flow turned
poleward again below 0.82~R${_\odot}$.

From all of the above it follows that in various studies, different authors
defined meridional flows of different directions and velocities.
In some studies significant temporal variations in the meridional flows were found,
though they were not always in complete agreement with each other.
The purpose of this study is to determine time-latitude variations in
meridional flows by analyzing  the dynamics of various strength
positive- and negative-polarity photospheric magnetic fields.
Section~\ref{S-data} describes the data used.
In Section~\ref{S-butterfly_WSO},  the time-latitude distributions of the large-scale
solar magnetic fields of positive and negative polarity in Cycles 21\,--\,24 are
investigated.
The time-latitude distributions of positive- and negative-polarity photospheric magnetic
fields using the data with high spatial resolution are analyzed in Section~\ref {S-butterfly_high_res}.
Variations in the magnitudes of the photospheric magnetic fields of
different strength are considered  in Section~\ref{S-mag}.
Mean latitudes and velocities of different type meridional circulations
are studied in Section~\ref{S-flow}.
The results are discussed  in Section~\ref{S-discussion}.
The main conclusions are listed in Section~\ref{S-conclusions}.

\section{Data}
  \label{S-data}

Synoptic magnetic field data from ground based WSO (Wilcox Solar Observatory,
\citep{Scherrer1977, Duvall1977, Hoeksema1983},
NSO KPVT (Kitt Peak Vacuum Telescope,
\cite{Livingston1976a,  Livingston1976b, Jones1992})
and SOLIS/VSM (Synoptic Optical Long-Term Investigations/Vector Stokes Magnetograph,
\cite{Keller2003}),
and space based SOHO/MDI (Solar and Heliospheric Observatory/Michelson
Doppler Imager, \cite{Scherrer1995})
and SDO/HMI (Solar Dynamics Observatory/Helioseismic
and Magnetic Imager, \cite{Scherrer2012, Schou2012}) observatories were used.
These observatories were chosen because they represent the most often
used data that overlap the greatest time interval of observations.
All synoptic maps were used in their original format
without any interpolation or changes.
Synoptic maps are maps of the magnetic field latitude-longitude distribution,
created  on the base of daily observations.
Each synoptic map span a full Carrington Rotation (CR, 1 CR = 27.2753 days).
The entire data set consists of 616 synoptic maps and covers CRs
1642\,--\,2258 (May 1976\,--\,May 2022).
Data on magnetic fields are given in a longitude
versus sine-latitude grid.

WSO provides the longest homogeneous series of observational data
of the low-resolution large-scale photospheric magnetic fields
\citep{Hoeksema1988} since 1976 without major updates of magnetograph.
The WSO synoptic maps represent the radial component of the photospheric
magnetic field, derived from observations of the line-of-sight field component
by assuming the field to be approximately radial.
WSO aperture size is 3 arc-min, which means
approximately 33 pixels in longitude in the equator.
WSO synoptic magnetic-field maps consist of $30$ data points in equal steps
of sine latitude from $70^{\circ}$S to $70^{\circ}$N.
Longitude is presented in $5^{\circ}$ intervals.
To convert WSO measured data to units of Gauss, a
factor of 1.85 is required \citep{Riley2014}.
"F-data" files, where missing data are interpolated, were applied.
WSO coronal magnetic field synoptic maps calculated from large-scale
photospheric fields with a potential field radial model with the
source-surface location at 2.5 R$_\odot$ \citep{Schatten1969,
Altschuler1969, Hoeksema1983} were also used.

Observations from KPVT (from 1976 to 2003, CRs 1625\,--\,2007)
and SOLIS (from 2003 to 2017, CRs 2007\,--\,2127)
were used to study the high-resolution photospheric magnetic fields.
At the KPVT, the pixel size of images was 1.0" from CR 1625 to CR 1853,
when a 512-channel Babcock type instrument was used,
and 1.14'' after CR 1855, when a CCD spectromagnetograph was used.
SOLIS's resolution was 1.125" until 2010, and  1.0" after that.
Every KPVT and SOLIS synoptic map consists of 360$\times$180 pixels
of magnetic field strength values in Gauss.
KPVT and SOLIS CR maps span 0$^\circ$\,--\,360$^\circ$ solar longitude
and from $90^{\circ}$S to $90^{\circ}$N sine solar latitude.

MDI instrument on board the
SOHO spacecraft was operational from 1996 to 2011.
There were data gaps in SOHO
for June\,--\,October 1998 and January\,--\,February 1999.
MDI  synoptic map consists of 3600$\times$1080 pixels.
MDI was succeeded by the HMI  in 2010, with a short overlapping period.
HMI provides magnetic field data with much higher
spatial and temporal resolutions and better quality.
The HMI synoptic maps have a size of 3600$\times$1440 pixels.

Synoptic data presented by WSO, KPVT, SOLIS, MDI, and HMI
are different \citep{Virtanen2016, Virtanen2017, Virtanen2019}.
The observatories use different instruments,
observation methods, data processing methods,
different spectral lines, and different technic
used for synoptic maps construction.
Their data have different spatial and spectral resolutions.
The instrumentation used at ground-based observatories
changed during the period concerned.
The method of polar magnetic field interpolation
in synoptic maps used in different
observatories also cause differences between the data sets.
Therefore the values of magnetic field strength
from different observatories differ significantly.
For comparison of magnetic field data obtained by
different instruments, the scaling coefficients were proposed
\citep{Pietarila2013, Riley2014}.

\section{Distributions and Meridional Flows of Large-Scale Magnetic Fields}
     \label{S-butterfly_WSO}

\begin{figure}
\centerline{\includegraphics[width=1\textwidth,clip=]{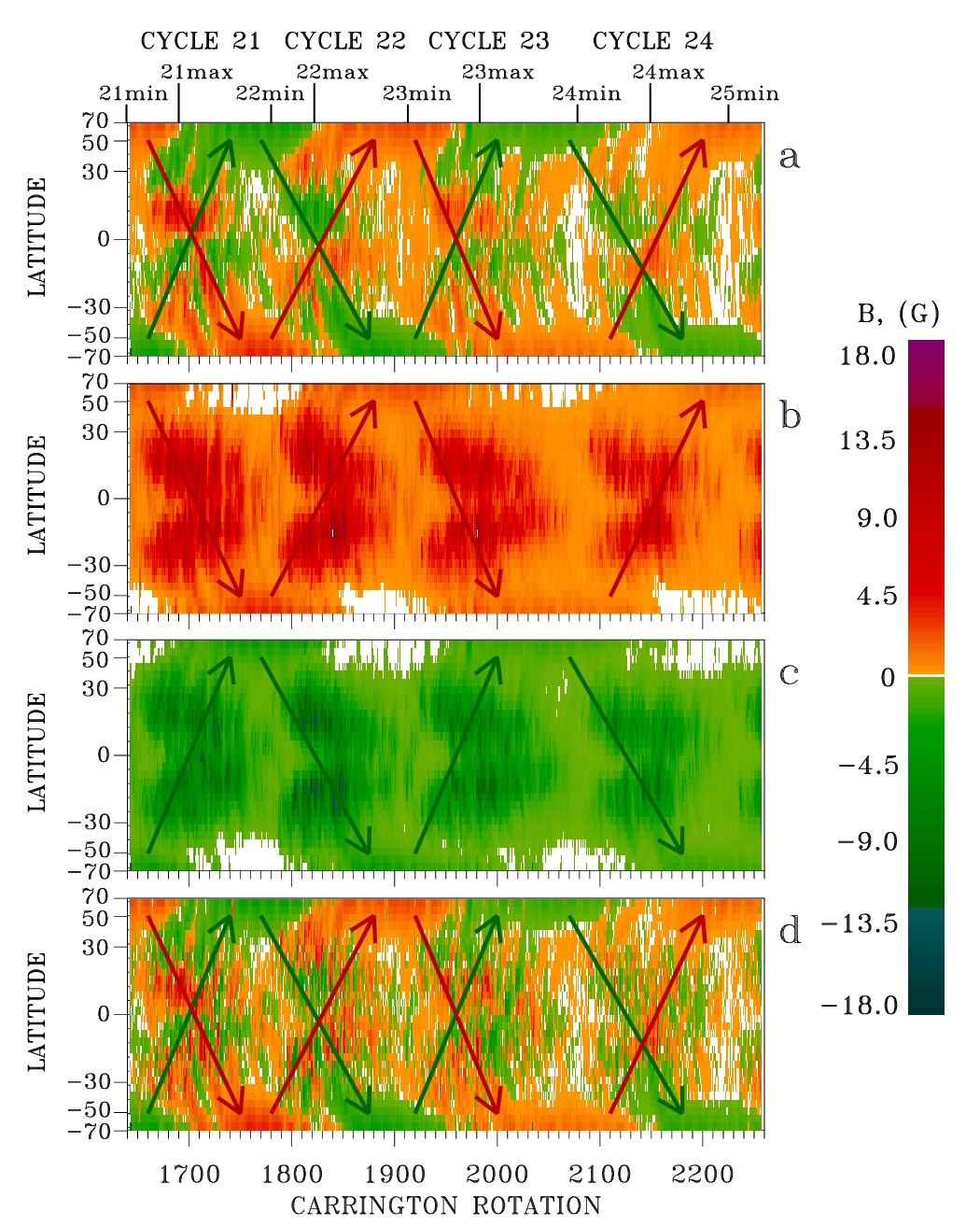}}
\caption{Distributions of large-scale photospheric magnetic fields (WSO).
(a) Butterfly diagram.
    Time-latitude distributions of longitude-averaged positive-polarity (b)
    and  negative-polarity (c) photospheric magnetic fields.
(d) Superposition of the distributions of positive- and negative-polarity
    magnetic fields shown in panels (b) and (c).
    Color indicates the field intensity.
 Red denotes the positive-polarity  and
 green denotes the negative-polarity magnetic fields.
 The maximum and minimum of each cycle are market at the top.}
 \label{batter_phot_wso}
 \end{figure}	

Butterfly diagrams are the main means of
studying solar meridional flows and creating various dynamo models.
When constructing a butterfly diagram, the longitude-average total
of the magnetic field for all latitudes for each CR is calculated.
The result is a diagram of the change in the average value of the magnetic
field in latitude with time.
Information about the changes in the longitudinal
distribution of magnetic fields in each CR is lost.
Figure~\ref{batter_phot_wso}(a) shows the time-latitude distribution of
longitudinally averaged large-scale photospheric magnetic fields,
the so-called butterfly diagram.
Red color indicates magnetic fields of positive polarity
and green that of negative polarity.
Different structures of magnetic fields of positive
and negative polarity were distinguished at different
phases of Solar Cycles 21\,--\,24.
A change in the sign of the magnetic field at the North and South
poles of the Sun is clearly seen.
The meridional circulation is believed to be directed
from the equator to the poles on the surface of the Sun.
But if similar diagrams are created separately for positive-
and negative-polarity magnetic fields, then their time-latitude
distributions will be completely different
(Figure~\ref{batter_phot_wso}(b~and~c)).
In these diagrams, low-strength positive- and negative-polarity magnetic fields
revealed wave-like pole-to-pole meridional flows.
The high-strength active region magnetic fields migrate from
high latitudes to the equator.
Red and green arrows show the directions of the
meridional wave flows on the photosphere
in each cycle (Figure~\ref{batter_phot_wso}(a\,--\,d)
as well as on the corresponding panels of all subsequent figures).
The meridional wave flows of positive- and negative-polarity
magnetic fields were antiphase in each cycle.
Each wave spanned a large range of latitudes in a CR.
They crossed the equator at the period of sunspot
maximum in each cycle.
During periods of solar activity minimum, they were at opposite poles.
The period of waves was equal to two solar cycles (approximately 22 years).
These waves are a manifestation of the meridional circulation
of the large-scale magnetic fields of positive and negative polarity
in each cycle.

Figure~\ref{batter_phot_wso}(d) presents the superposition of the
distributions of positive- and negative-polarity
magnetic fields shown in Figure~\ref{batter_phot_wso}(b~and~c).
Comparison of diagrams in Figure~\ref{batter_phot_wso}(a~and~d)
shows that they are almost the same.
Some small differences  are due to
the fact that the average value of total magnetic fields for
each CR is not equal to the sum of the averages
of positive- and negative-polarity magnetic fields for the same CR.

\begin{figure}
\centerline{\includegraphics[width=1\textwidth,clip=]{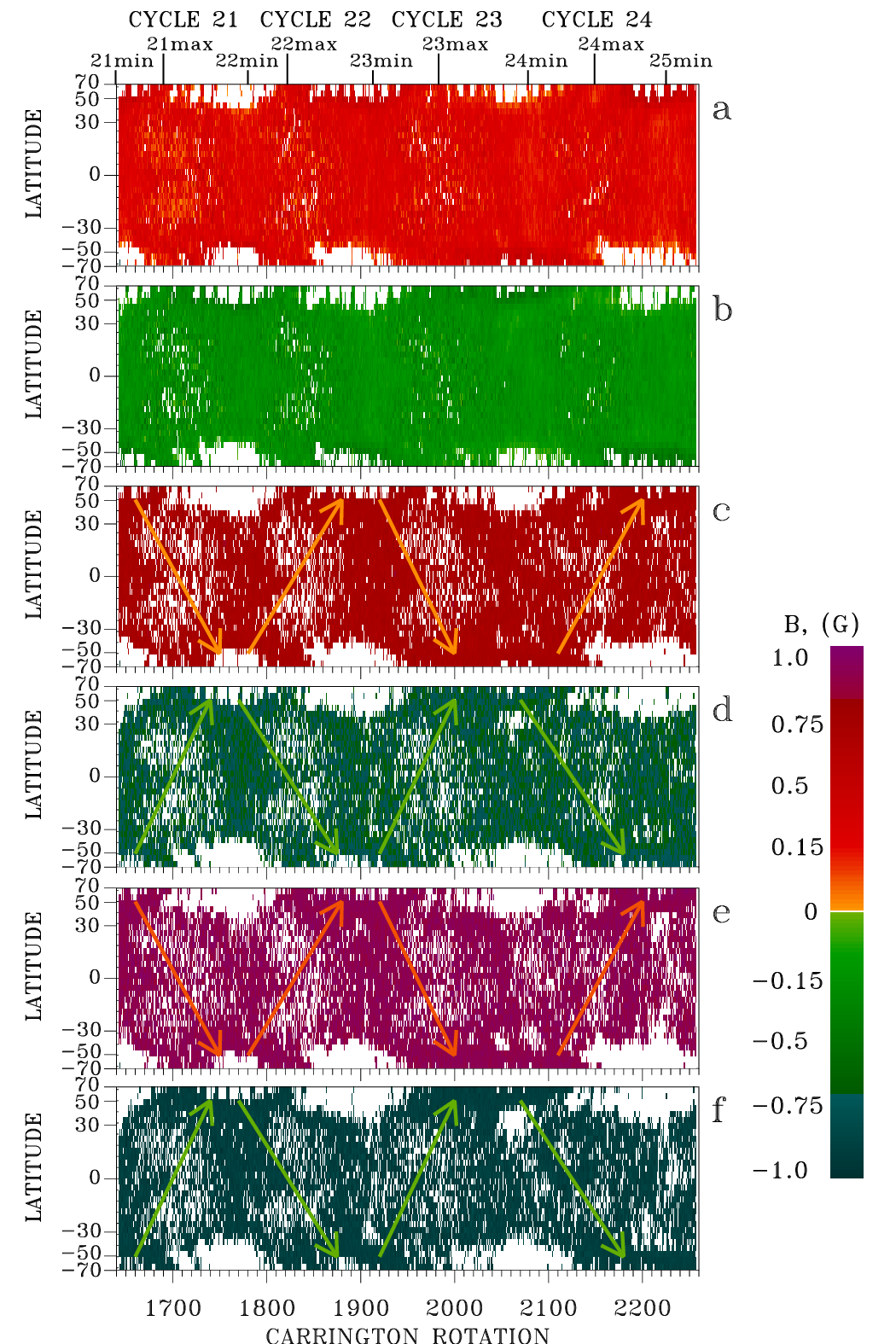}}
\caption{Time-latitude distributions of large-scale positive- and negative-polarity
          low-strength photospheric magnetic fields in different ranges (WSO).
 		(a) $ 0<B\le0.6$~G;
 		(b)  $-0.6\le B<0$~G;
 		(c) $0.6<B\le0.8$~G;
 		(d)  $-0.8\le B<-0.6$~G;
 		(e) $0.8<B\le1$~G;
 		(f) $ -1\le B<-0.8$~G.
 		Designations are the same as in Figure~\ref{batter_phot_wso}.}
 	\label{batter_phot_wso_low}
 \end{figure}

\begin{figure}
\centerline{\includegraphics[width=1\textwidth,clip=]{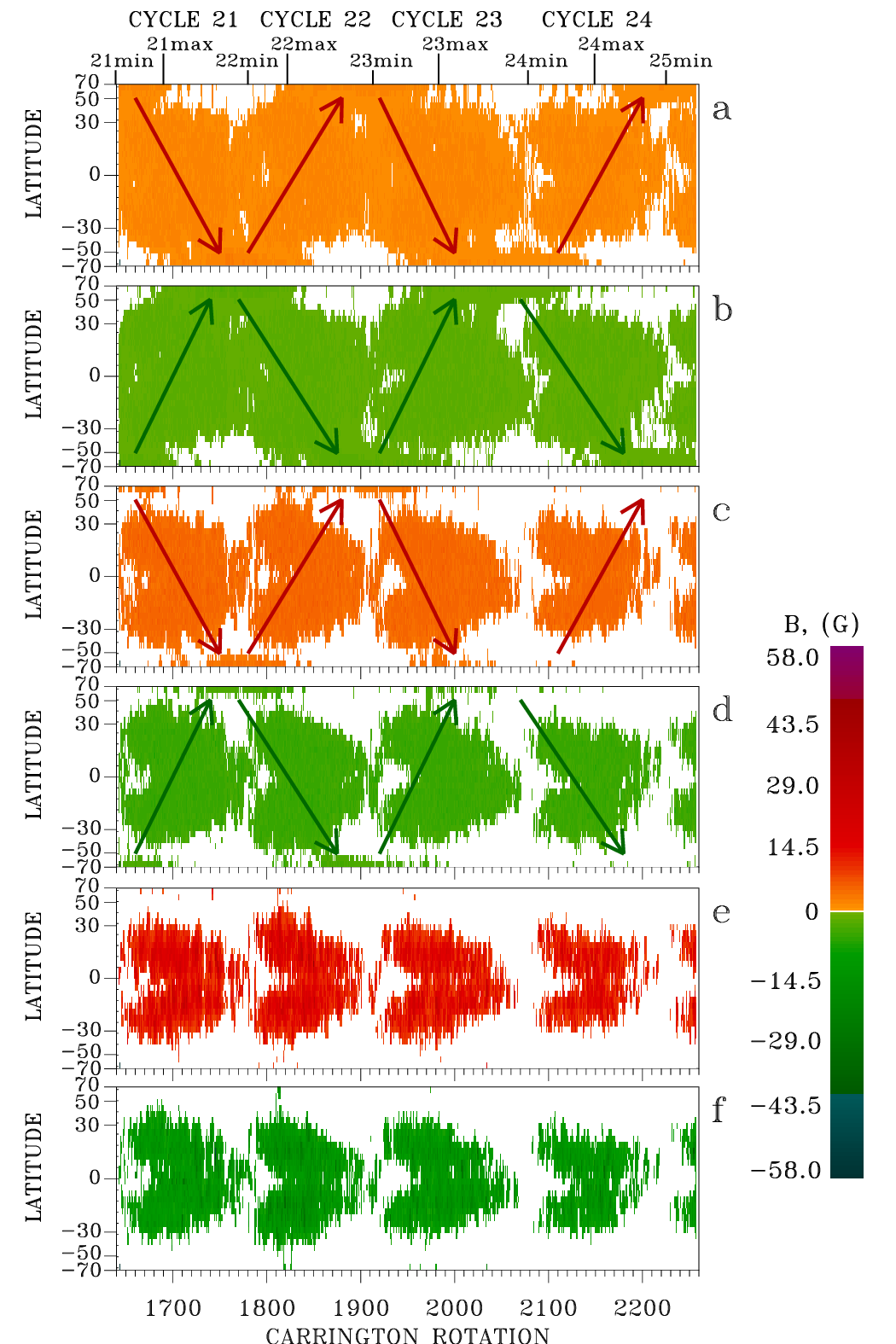}}
\caption{Time-latitude distributions of large-scale positive- and negative-polarity
            high-strength photospheric magnetic fields in different ranges (WSO).
 		(a)  $1<B\le 3$~G;
 		(b)  $ -3\le B<-1$~G;
 		(c)  $3<B\le 7$~G;
 		(d)  $-7\le B<3$~G;
 		(e)  $B>7$~G;
 		(f)  $B< -7$~G.
 		Designations are the same as in Figure~\ref{batter_phot_wso}.}
 	\label{batter_phot_wso_high}
 \end{figure}

\begin{figure}
\centerline{\includegraphics[width=1\textwidth,clip=]{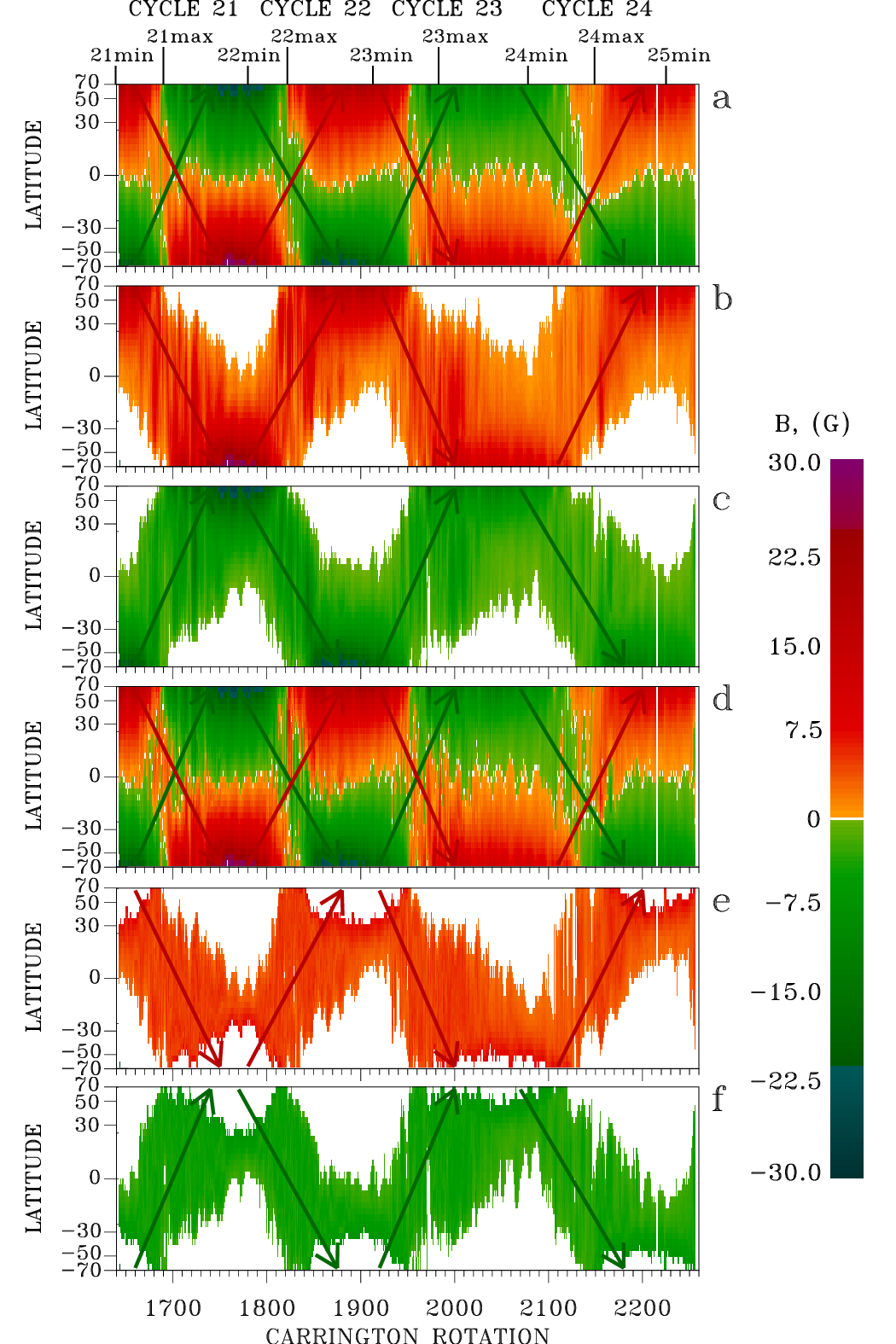}}
\caption{Time-latitude distributions of large-scale magnetic fields
                   calculated at the source surface (2.5 R$_\odot$, WSO).
 (a) Butterfly diagram.
     Time-latitude distributions of longitude-averaged positive-polarity (b)
     and  negative-polarity (c) magnetic fields.
 (d) Superposition of the distributions of positive- and negative-polarity
     magnetic fields shown in panels (b) and (c).
     Time-latitude distributions of positive-polarity (e) and negative-polarity (f)
     magnetic fields in the range of $2<\mid B \mid\le 7$~G.
      Designations are the same as in Figure~\ref{batter_phot_wso}.}
 	\label{batter_r25_wso}
 \end{figure}

Meridional flows of large-scale positive- and negative-polarity magnetic fields
with different  strength are shown in
Figure~\ref{batter_phot_wso_low} for magnetic fields $ 0<\mid B \mid\le 0.6$~G
and in Figure~\ref{batter_phot_wso_high} for magnetic fields $\mid B \mid\ge$1~G.
Note that the color bars in Figures~\ref{batter_phot_wso}\,--\,\ref{batter_phot_wso_high}
are different.
Magnetic field variations in Figure~\ref{batter_phot_wso_low}  reflect
the dynamics of low- and medium-strength magnetic fields.
The distributions of magnetic fields in the range of $0<\mid B \mid\le 0.6$~G
were almost the same.
Both positive- and negative-polarity magnetic fields were distributed
evenly from the South to the North pole.
Pole-to-pole meridional waves of positive- and negative-polarity
magnetic fields appear from approximately $\mid B \mid\ \approx 0.6$~G.
They were antiphase and antisymmetrical with respect to the equator.
At the solar cycle minima the wave-like meridional flows
were shifted to high latitudes in opposite hemispheres.
The change in their meridional flow direction occurred systematically
in the rising and ascending phases of each solar cycle.
During periods of solar maximum, the meridional flows
of positive and negative polarity got closer
and approached the equator. Then they crossed the equator
and continued to migrate to the opposite poles.
From Figures~\ref{batter_phot_wso} and
\ref{batter_phot_wso_low} it follows that these meridional flows
determine the process of solar polar field reversal.

As the magnetic field strength increases, the wave-like
meridional flows disappear.
The active region (high-strength) magnetic fields remain only.
Figure~\ref{batter_phot_wso_high} presents the
time-latitude distributions of large-scale longitude-averaged
high-strength positive- and  negative-polarity photospheric
magnetic fields in different ranges.
The magnetic fields of active regions began to dominate
initially in low Solar Cycles 23 and 24.
In high Solar Cycles 21 and 22, the meridional wave flows were
observed up to  $\mid B \mid\approx3$~G.
Whereas in low Solar Cycles 23 and 24, these meridional flows
ceased to be observed at lower magnetic fields,
$\mid B \mid\approx1$~G.
Meridional flows of active region magnetic fields
both positive and negative (leading and following sunspot) polarity
were symmetric with respect to the equator and they
migrated from high to low latitudes in the Northern
and Southern hemispheres in all ranges.

Figure~\ref{batter_r25_wso} shows similar magnetic field meridional
circulation diagrams for coronal magnetic fields calculated from large-scale
photospheric fields (WSO) with a potential radial field model with the
source-surface location at 2.5~R$_\odot$.
The same arrows are plotted on all panels,
as in Figure~\ref{batter_phot_wso}.
It should be emphasized that the magnetic fields of active
regions are absent from these diagrams, since they are believed
to be mainly confined below the source surface.
Therefore large-scale magnetic field dynamics is only presented.
The dynamics of magnetic fields on the source surface
reflects the cycle variations of the solar global magnetic field.
Figure~\ref{batter_r25_wso}(a) presents the total distribution of all
magnetic fields.
In Figure~\ref{batter_r25_wso}(b), the time-latitude distribution
of longitude averaged positive-polarity magnetic fields and in
Figure~\ref{batter_r25_wso}(c), that of the
negative-polarity are presented.
Superposition of positive- and negative-polarity
magnetic field distributions is shown in Figure~\ref{batter_r25_wso}(d).
Latitudinal distributions of positive- and negative-polarity
magnetic fields in the range of $2<\mid B \mid \le7$~G
(without polar fields) are presented in Figure~\ref{batter_r25_wso}(e~and~f).
The wave-like, pole-to-pole, antiphase meridional flows of the positive-
and negative-polarity magnetic fields stand out even more clearly
in Figure~\ref{batter_r25_wso}.

Thus, three types of time-latitude distributions and meridional circulations
of large-scale magnetic fields, depending on the field strength, have been identified.
The first one is the distribution of low-strength positive- and
negative-polarity magnetic fields ($0<\mid B \mid \le $0.6~G).
They distributed uniformly across the solar disk and
symmetrically with respect to the equator.
The second type is the wave-like, antiphase, pole-to-pole meridional circulation
of medium-strength positive- and negative-polarity magnetic fields
in the range of  $0.6<\mid B \mid \le 3.0$~G.
The third type is the well known distribution of strong
magnetic fields of active regions that show meridional flows
from high latitudes toward the equator for both positive-
and negative-polarity (leading and following sunspot polarity) magnetic fields.
From Figures~\ref{batter_phot_wso_low} and
\ref{batter_phot_wso_high} it follows that
the time-latitude distributions of low- and high-strength magnetic field
are very different from each other.
This indicates the independent formation of low-strength
magnetic fields and argues in favor of the small-scale dynamo theory.
The meridional circulation of the medium-strength photospheric magnetic fields
is also very different from that of active regions.
This suggests that they have different sources
of their generation and cycle evolution.
The wavy pole-to-pole meridional circulation of
medium-strength magnetic fields indicates that it is these magnetic fields
that determine the process of solar polar field reversal,
and not the magnetic fields of active regions.

The butterfly diagram is the result of a superposition
of magnetic fields of different strengths and their meridional circulations.
The various details and structures in the butterfly diagram,
for example poleward surges, are the result of the domination
of magnetic field polarity of one of the meridional circulation type.

\section{Distributions and Meridional Flows of High Spatial Resolution Magnetic Fields}
    \label{S-butterfly_high_res}

\begin{figure}
\centerline{\includegraphics[width=1\textwidth,clip=]{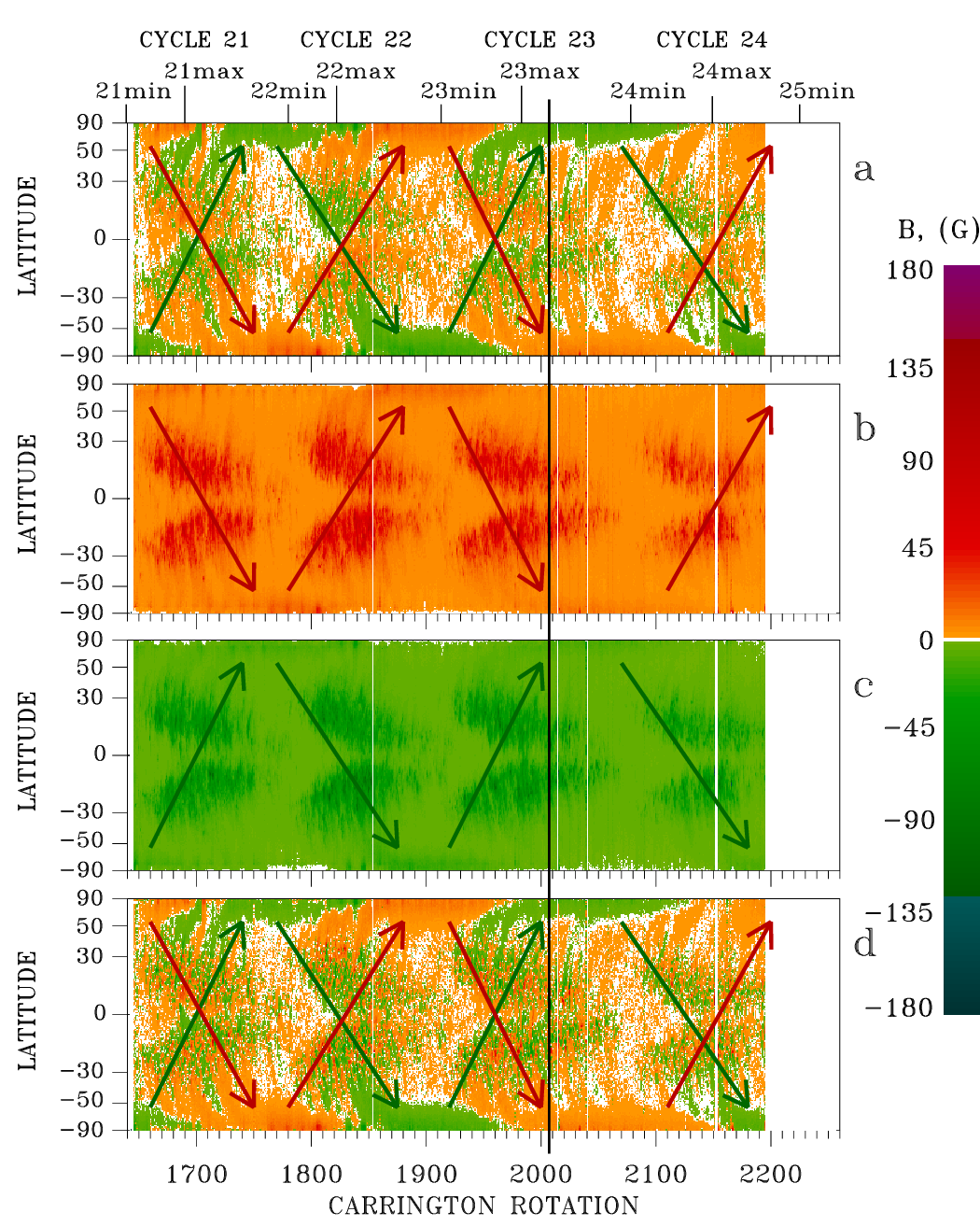}}
\caption{Distributions of high-resolution photospheric magnetic fields
     from KPVT (CRs 1625\,--\,2007) and SOLIS (CRs 2008\,--\,2196) synoptic maps.
(a) Butterfly diagram.
    Time-latitude distributions of longitude-averaged positive-polarity (b) and
    negative-polarity (c) magnetic fields.
(d) Superposition of the distributions of positive- and negative-polarity
       magnetic fields shown in panels (b) and (c).
    Designations are the same as in Figure~\ref{batter_phot_wso}.
    Black vertical line marks the transition from KPVT to SOLIS in CR 2007.}
	\label{batter_kp}
\end{figure}

\begin{figure}
\centerline{\includegraphics[width=1\textwidth,clip=]{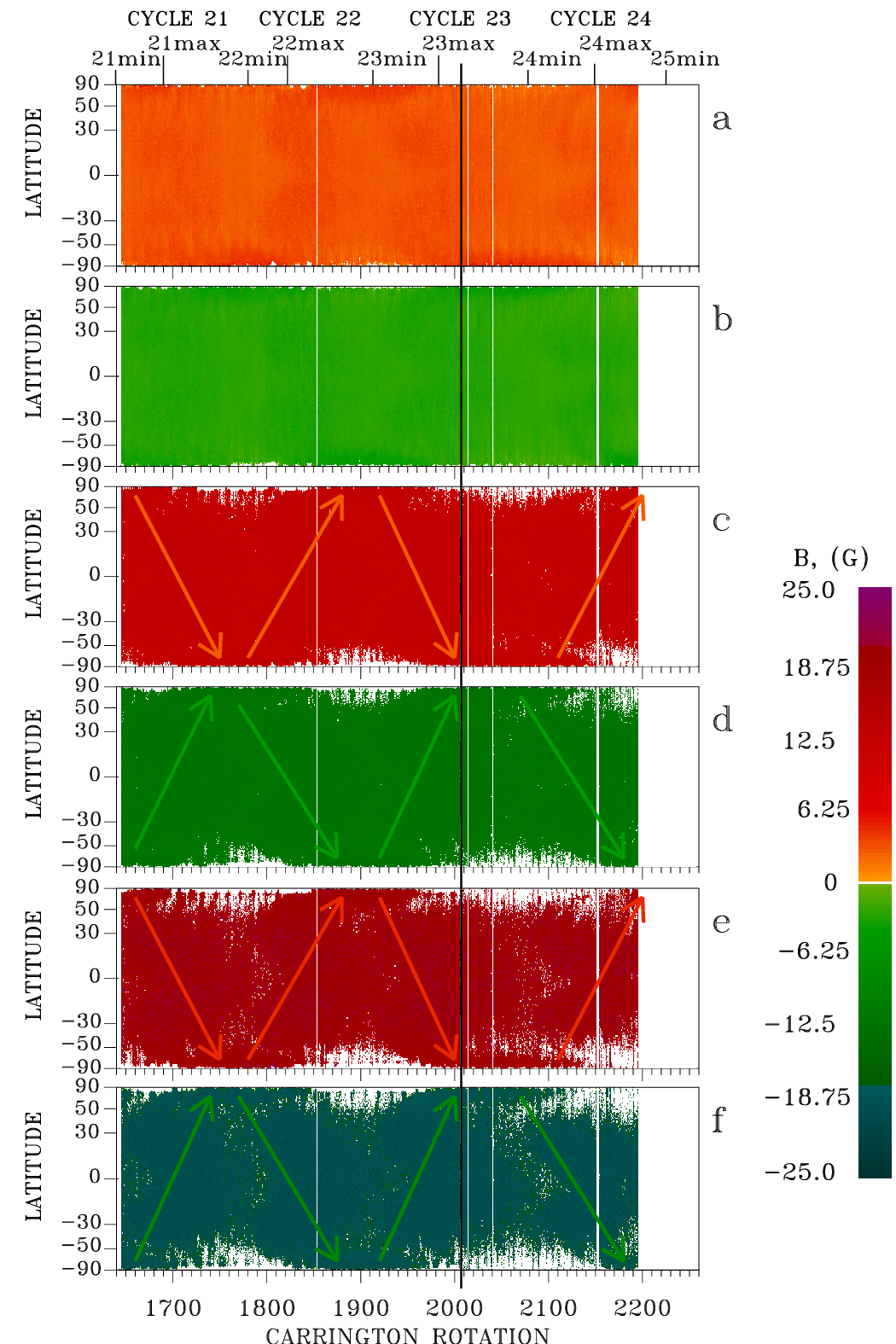}}
\caption{Time-latitude distributions of high-resolution longitude-averaged low-strength
           and medium-strength positive- and negative-polarity photospheric magnetic fields
         in different ranges (KPVT, CRs 1625\,--\,2007 and SOLIS, CRs 2008\,--\,2196).
        (a)  $0<B\le 7$~G;
        (b)  $-7\le B<0$~G;
        (c)  $10<B\le 15$~G;
        (d)  $-15\le B<-10$~G;
        (e)  $15<B\le 25$~G;
        (f)  $-25\le B<-15$~G.
	Designations are the same as in Figure~\ref{batter_kp}.}
	\label{batter_kp_low}
\end{figure}

\begin{figure}
\centerline{\includegraphics[width=1\textwidth,clip=]{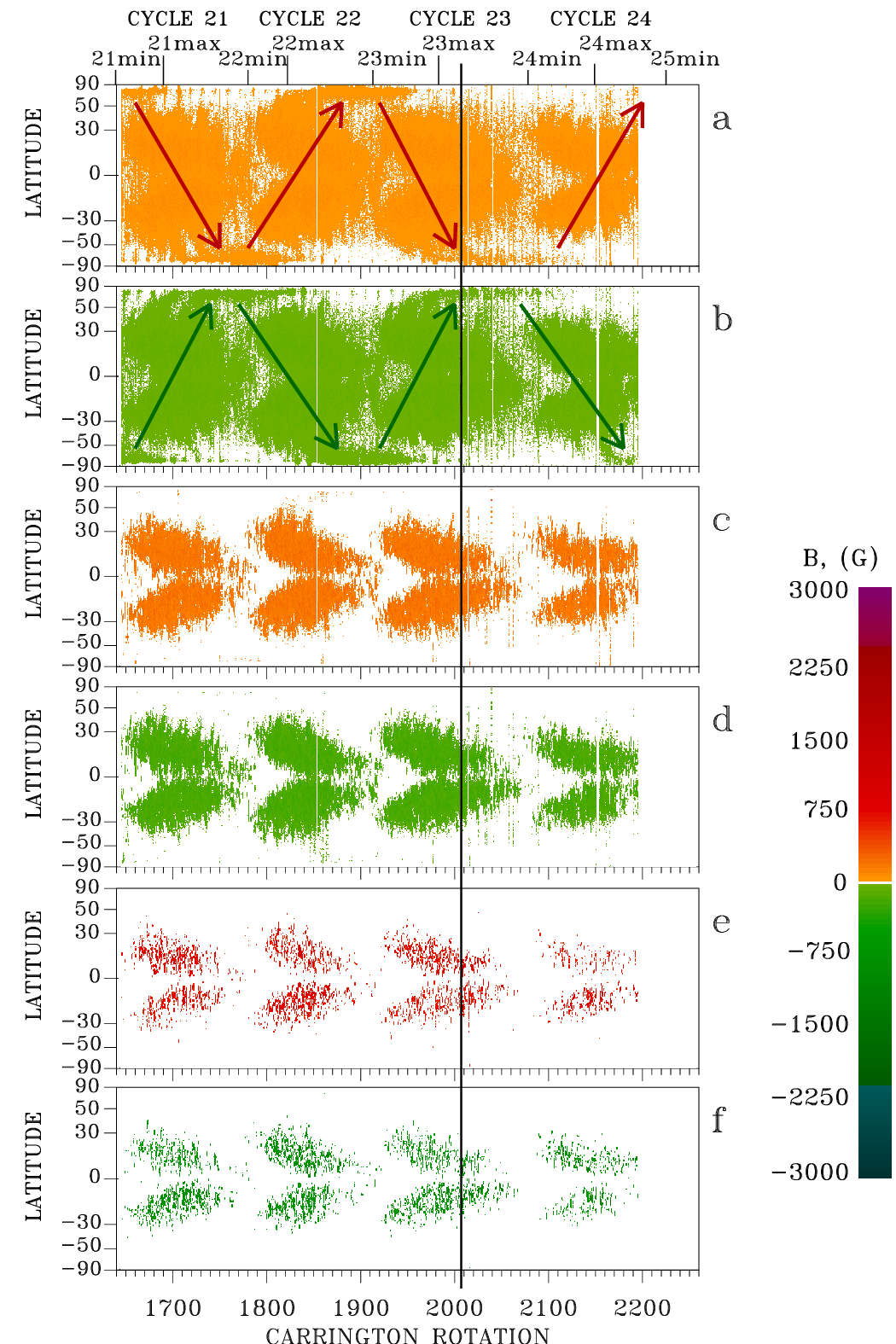}}
\caption{Time-latitude distributions of high-resolution longitude-averaged high-strength
         positive- and negative-polarity photospheric magnetic fields in different ranges
         (KPVT, CRs 1625\,--\,2007 and SOLIS, CRs 2008\,--\,2196).
	    (a)  $25<B\le 100$~G;
        (b)  $-100\le B<-25$~G;
        (c)  $100<B\le 500$~G;
        (d)  $-500\le B<-100$~G;
        (e)  $B> 500$~G;
        (f)  $B< -500$~G.
	Designations are the same as in Figure~\ref{batter_kp}.}
	\label{batter_kp_high}
\end{figure}

\begin{figure}
\centerline{\includegraphics[width=1\textwidth,clip=]{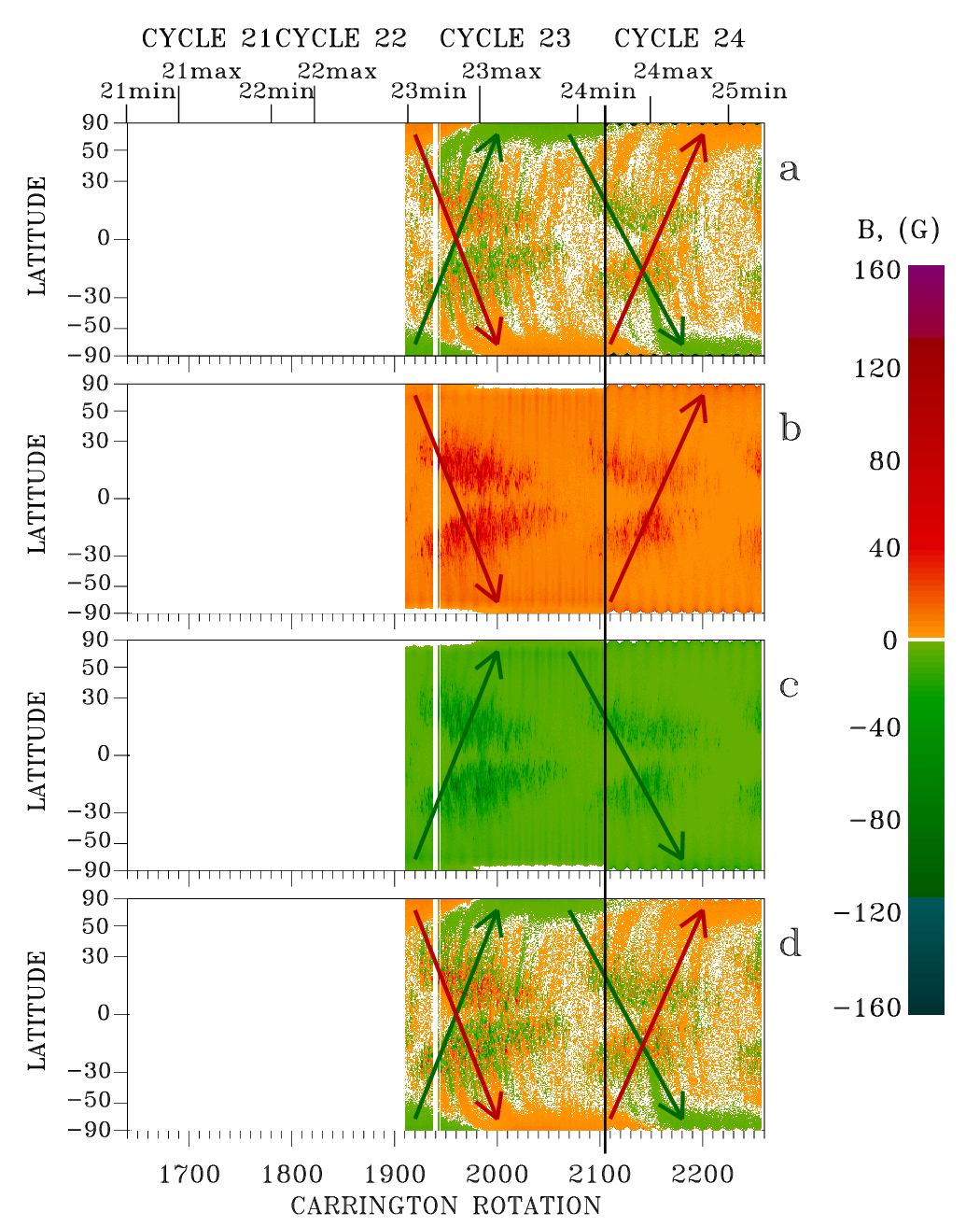}}
\caption{Distributions of high-resolution photospheric magnetic fields from
	 SOHO/MDI (CRs 1911\,--\,2104) and SDO/HMI (CRs 2105\,--\,2267) synoptic maps.
(a) Butterfly diagram.
    Time-latitude distributions of longitude-averaged positive-polarity (b)
    and negative-polarity (c) magnetic fields.
(d) Superposition of the distributions of positive- and negative-polarity
       magnetic fields shown in panels (b) and (c).
    Black vertical line marks the transition from
    SOHO/MDI to SDO/HMI in CR 2105.
    Designations are the same as in Figure~\ref{batter_phot_wso}.}
\label{batter_sh}
\end{figure}

\begin{figure}
\centerline{\includegraphics[width=1\textwidth,clip=]{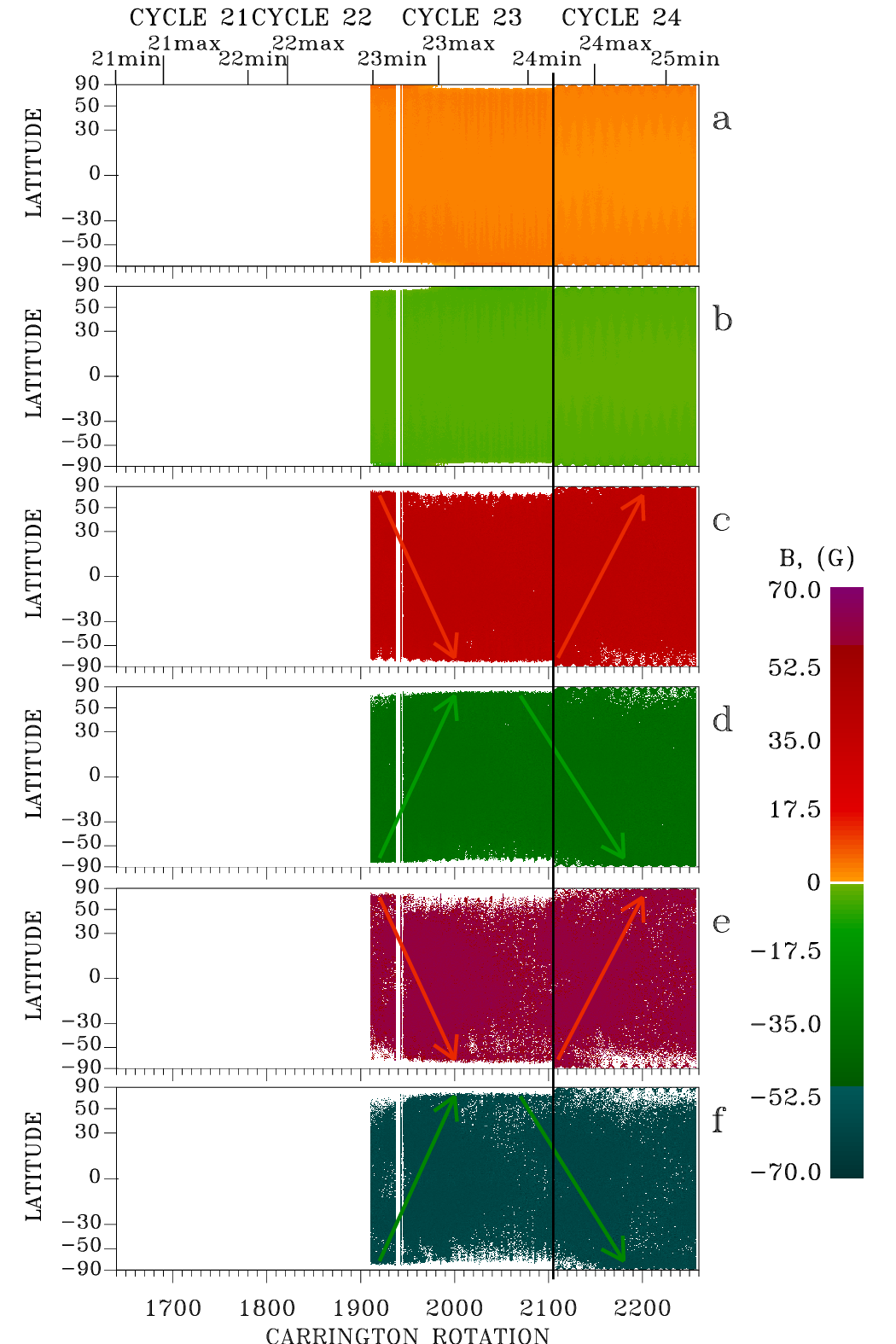}}
\caption{Time-latitude distributions of high-resolution longitude-averaged low-strength
      and medium-strength positive- and negative-polarity photospheric magnetic fields
 		(SOHO/MDI, CRs 1911\,--\,2104 and SDO/HMI, CRs 2105\,--\,2267).
	    (a)  $0<B\le 7$~G;
        (b)  $-7\le B<0$~G;
        (c)  $30<B\le 50$~G;
        (d)  $-50\le B<-30$~G;
        (e)  $50<B\le 70$~G;
        (f)  $-70\le B<-50$~G.
	Designations are the same as in Figure~\ref{batter_sh}.}
 	\label{batter_sh_low}
\end{figure}

\begin{figure}
\centerline{\includegraphics[width=1\textwidth,clip=]{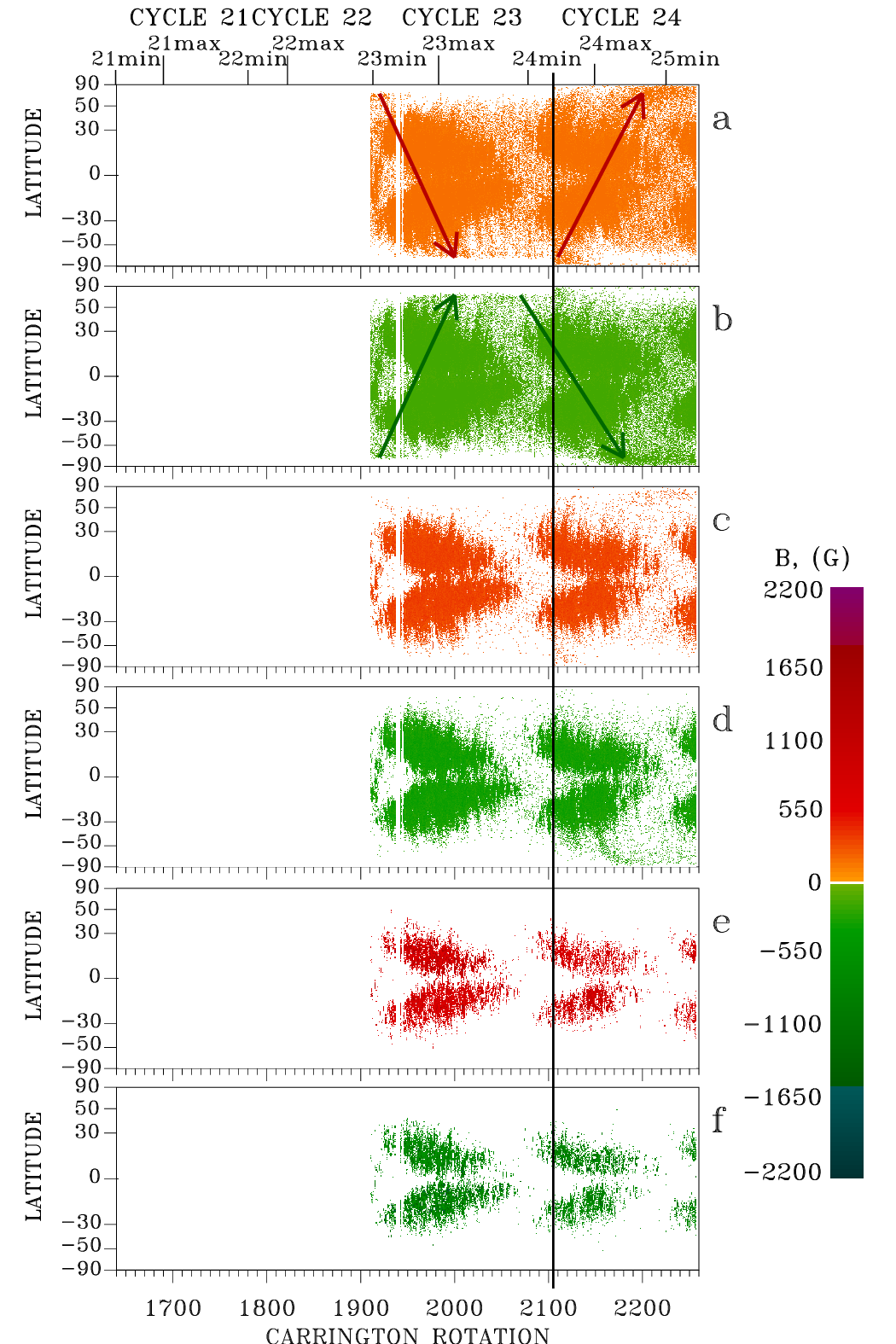}}
\caption{Time-latitude distributions of high-resolution longitude-averaged high-strength
         positive- and negative-polarity photospheric magnetic fields in different ranges
 		(SOHO/MDI, CRs 1911\,--\,2104 and  SDO/HMI, CRs 2105\,--\,2267).
	    (a)  $100<B\le 200$~G;
        (b)  $-200\le B<-100$~G;
        (c)  $200<B\le 500$~G;
        (d)  $-500\le B<-200$~G;
        (e)  $B>500$~G;
        (f)  $B<-500$~G.
       Designations are the same as in Figure~\ref{batter_sh}.}
 	\label{batter_sh_high}
\end{figure}

The resolution of WSO magnetic fields is 3 arc-min.
Therefore, WSO magnetic-field data show the dynamics
of low-resolution large-scale magnetic fields.
To investigate the distribution and meridional circulation
of high-resolution magnetic fields,
synoptic data from ground based KPVT and SOLIS  and space SOHO/MDI
and SDO/HMI observatories were used.
Figure~\ref{batter_kp} shows the time-latitude distributions of
longitude-averaged positive- and negative-polarity magnetic fields
at high spatial resolution (KPVT, CRs 1625\,--\,2007 and SOLIS, CRs 2008\,--\,2196).
Figure~\ref{batter_kp_low} shows that in the range of $0<\mid B\mid\le 25$~G,
and Figure~\ref{batter_kp_high} shows that for $\mid B \mid>25$~G.
Figures~\ref{batter_sh}\,--\,\ref{batter_sh_high}
show similar time-latitude distributions constructed
from the synoptic magnetic maps of SOHO/MDI (CRs~1911\,--\,2104)
and SDO/HMI (CRs~2105\,--\,2267).
HMI maps (3600$\times$1440 pixels) were recalculated
to MDI scale (3600$\times$1080 pixels).
To convert HMI magnetic field data to that of MDI,
the conversion factor of 1.01 \citep{Riley2014} was used.

Unlike WSO, the fine structure of the magnetic field distribution appears more
clearly in the diagrams Figures~\ref{batter_kp}~and~\ref{batter_sh}.
Comparison of Figures~\ref{batter_phot_wso}~\,--\,~\ref{batter_sh_high}
shows that different data show a very similar distributions and
meridional flows in similar magnetic field ranges, but
the intensity of the magnetic fields
greatly differs between the data sets.
It should be noted that magnetic field distributions
were analyzed on the basis of the synoptic maps of
different independent observatories.
Despite this, the results obtained using instruments
with low and high spatial resolution are the same.
Three types of magnetic field distributions
and meridional circulations depending on the magnetic
field strength were also identified.
The first type are low-strength magnetic fields ($0<\mid B \mid \le 7$~G,
KPVT-SOLIS and MDI-HMI).
They were almost cycle-independent and distributed evenly over the solar disk.
The second type includes medium-strength magnetic fields
($10<\mid B \mid\le 25$~G, KPVT-SOLIS and $30<\mid B \mid\le 70$~G, MDI-HMI)
High-resolution medium-strength magnetic fields
also revealed  wave-like, antiphase, pole-to-pole
meridional circulation with a period of $\approx$22 years.
They transport the new polarity magnetic field to the poles.
The third type consists of strong magnetic fields
($\mid B \mid > 100$~G, KPVT-SOLIS and $\mid B \mid > 200$~G, MDI-HMI).
High-strength magnetic fields show meridional
circulation of active regions from high  to low latitudes
in the Northern and Southern hemispheres.
Neither the leading nor the following polarity sunspot magnetic fields
migrate to the poles in any of the magnetic field ranges.

From Figures~\ref{batter_kp_high}(a~and~b)~and~
\ref{batter_sh_high}(a~and~b) it follows that some
high-latitude active region magnetic fields, whose polarity
coincided with the polarity of the second type meridional circulation waves,
i.e. medium-strength magnetic field flows, indicated by arrows
in Figures~\ref{batter_kp_high}(a~and~b)~and~\ref{batter_sh_high}(a~and~b),
were picked up by the second type pole-to-pole meridional circulation
waves and transported to the poles.
Thus some active region magnetic fields participate in
the solar pole field reversals, but they are not the main source
of the new polarity magnetic fields at the solar poles.

\section{Variations in the Magnetic Field Magnitudes of Different Type Meridional Circulations}
     \label{S-mag}

\begin{figure}
\centerline{\includegraphics[width=1\textwidth,clip=]{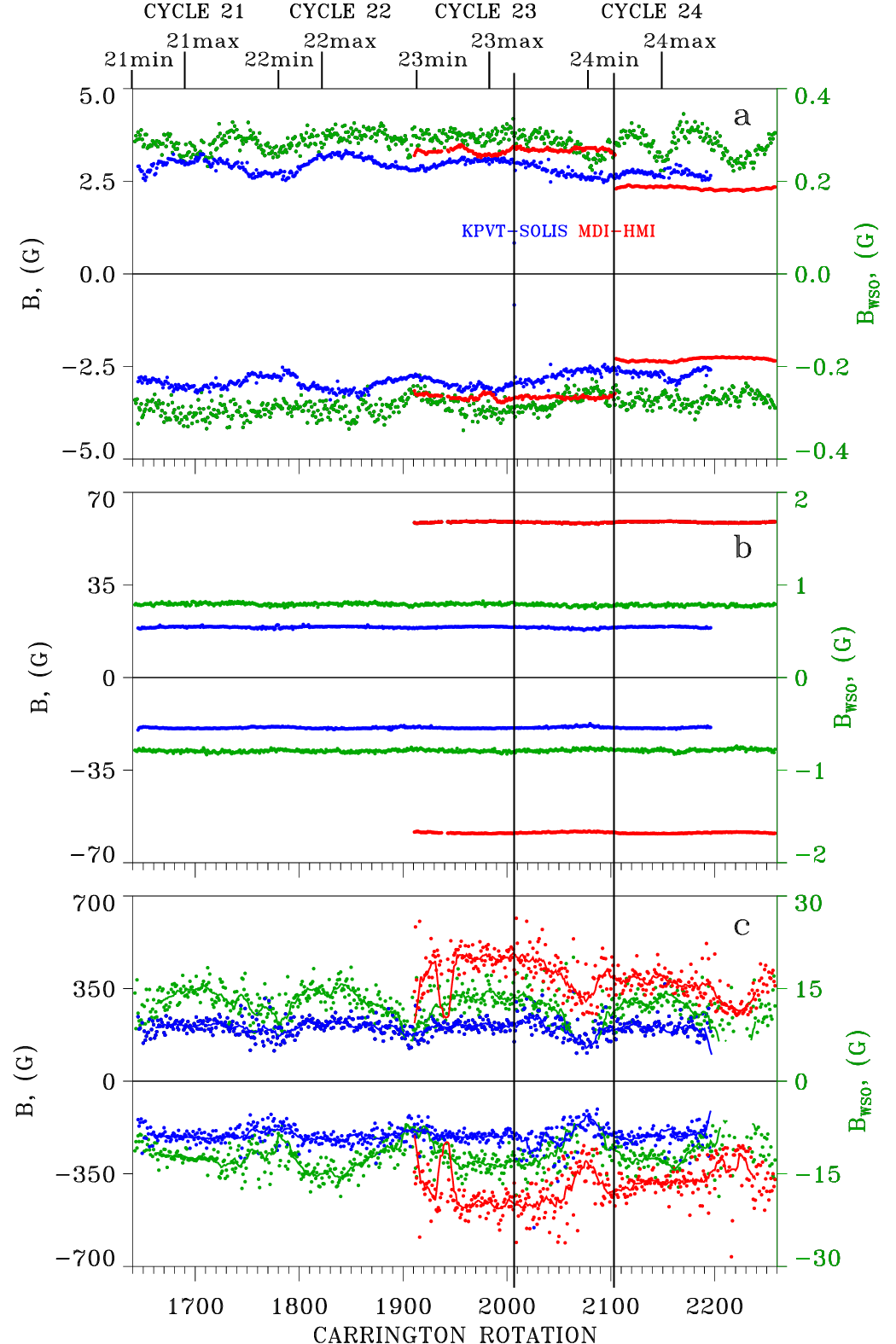}}
\caption{Variations in the mean magnetic field  strength.
		(a)   $ 0 <\mid B \mid\le 0.6 $~G (WSO);
	        	$ 0 <\mid B \mid\le 7 $~G (KPVT-SOLIS and MDI-HMI);   	
		(b)    $ 0 .6<\mid B \mid\le 1.0 $~G (WSO);
	        	 $ 10<\mid B \mid\le 25 $~G (KPVT-SOLIS);
		        $ 30 <\mid B \mid\le 70 $~G (MDI-HMI);
		(c)   $ \mid B \mid >7 $~G (WSO);
		        $ \mid B \mid >100 $~G (KPVT-SOLIS);
                $ \mid B \mid >200 $~G (MDI-HMI).
   Green denotes WSO data (right $y$ axes),
   blue  denotes KPVT-SOLIS data, and
   red  denotes MDI-HMI data.
Dots represent CR-averaged data and thick lines represent seven-CR-averaged data.
Solar cycles maxima and minima are marked at the top.
The black vertical lines mark  the transition
from KPVT to SOLIS and from MDI to HMI.}
	\label{mag}
\end{figure}

Figure~\ref{mag} shows the variations in the mean
strength of magnetic fields in each CR for the selected
magnetic-field ranges  according to WSO (green),
KPVT-SOLIS (blue), and MDI-HMI (red) data.
Variations in the mean of low-strength magnetic fields
(the first type) are shown in Figure~\ref{mag}(a)
Variations in WSO ($0<\mid B \mid \le 0.6$~G) data were
insignificant in amplitude and did not coincide with
the solar cycle variations of active region magnetic fields.
They are minimal at cycles maxima.
The mean magnetic field values of KPVT-SOLIS data ($0<\mid B\mid\le 7$~G)
increased slightly toward the maximum and decreased
toward the minimum of each solar cycle.
MDI-HMI low-strength magnetic fields  did not reveal solar cycle variations
at all.

The mean magnetic field values of the second type meridional circulation, i.e.
the medium-strength wave-like, pole-to-pole
positive- and negative-polarity magnetic field flows,
are presented in Figure~\ref{mag}(b).
The selected magnetic field ranges correspond to
$ 0.6<\mid B \mid\le 1.0 $~G  for WSO,
$ 15<\mid B \mid\le 25 $~G  for KPVT-SOLIS, and
$ 50 <\mid B \mid\le 70 $~G for MDI-HMI data.
For the second type meridional circulation, the magnitudes
of magnetic fields remained approximately at the same level
at all solar cycle phases during Solar Cycles 21\,--\,24.

In Figure~\ref{mag}(c) the variations in the mean magnetic field of
the third type meridional circulation (meridional flows of high-strength
positive- and negative-polarity active region magnetic fields) are shown.
The selected magnetic field ranges correspond to
$ \mid B \mid >7 $~G for WSO,
$ \mid B \mid >100 $ for KPVT-SOLIS, and
$ \mid B \mid >200 $~G for  and MDI-HMI.
The solar cycle variations are clearly visible.
The mean values of magnetic fields in these ranges
increased to the maximum and decreased to
the minimum of solar activity in each cycle.

The mean of magnetic fields  calculated on the base of
high-resolution synoptic maps, turn out to be low, since weak
magnetic fields occupy larger areas in the synoptic maps and their
contribution to the total magnetic field is higher.
It should be noted that the MDI and HMI data in the
range of $0 <\mid B \mid\le7.0$~G did not agree well
with each other despite the using of conversion coefficient.
The values of magnetic fields according
to the HMI data were much lower than those according to the MDI data.
There is no such difference in other magnetic field ranges.
This may indicate the nonlinearity of the relationship between
the MDI and HMI magnetic field data.

\section{Mean Latitudes and Velocities of Different Type Meridional Circulations}
    \label{S-flow}

\begin{figure}
\centerline{\includegraphics[width=1\textwidth,clip=]{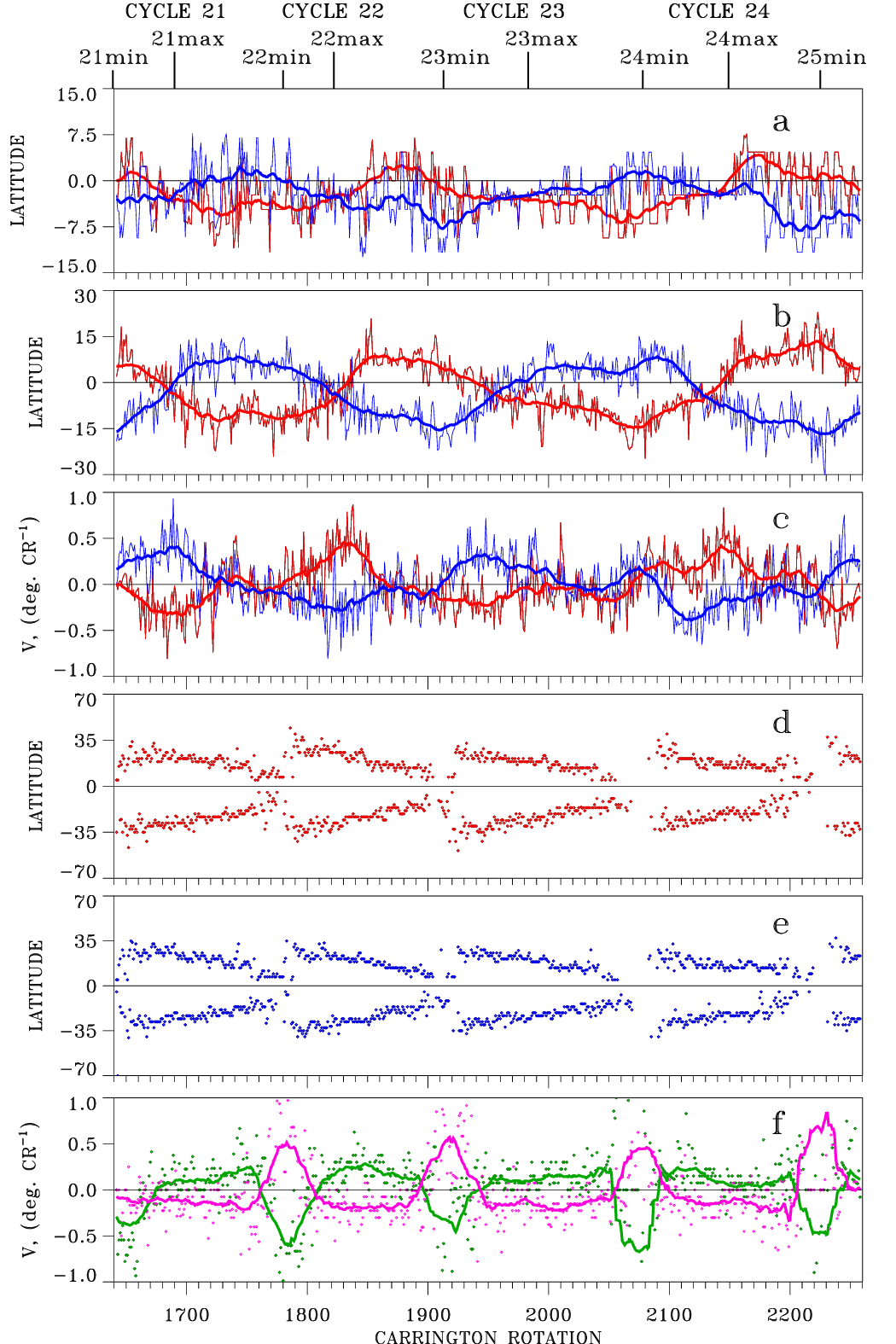}}
\caption{Mean latitudes and velocities of different type
         meridional flows (WSO).
         Variations in the mean latitudes of low-strength positive- and negative-polarity
         magnetic fields in the range of $0<\mid B \mid\le 0.6$~G  (a).
   (b)   Variations in the mean latitudes of medium-strength positive-
         and negative-polarity magnetic fields in the range of $0.8<\mid B \mid\le 1.0$~G.
   (c) Velocities of medium-strength positive-polarity (red) and negative-polarity (blue)
         magnetic field flows.
         Variations in the mean latitudes of high-strength positive-polarity (d) and
         negative-polarity (e) magnetic fields in the range of $\mid B \mid>7$~G.
   (f)    Velocities of the high-strength positive-polarity  magnetic field flows
           in the North (green) and  South (lilac) hemispheres.
In (a)\,--\,(e) red indicates magnetic fields of positive polarity,
and blue indicates that of negative polarity.
Thin lines show the values averaged for each CR.
Thick lines indicate the values averaged over 31 CRs.}
	\label{lat_wso}
\end{figure}

\begin{figure}
\centerline{\includegraphics[width=1\textwidth,clip=]{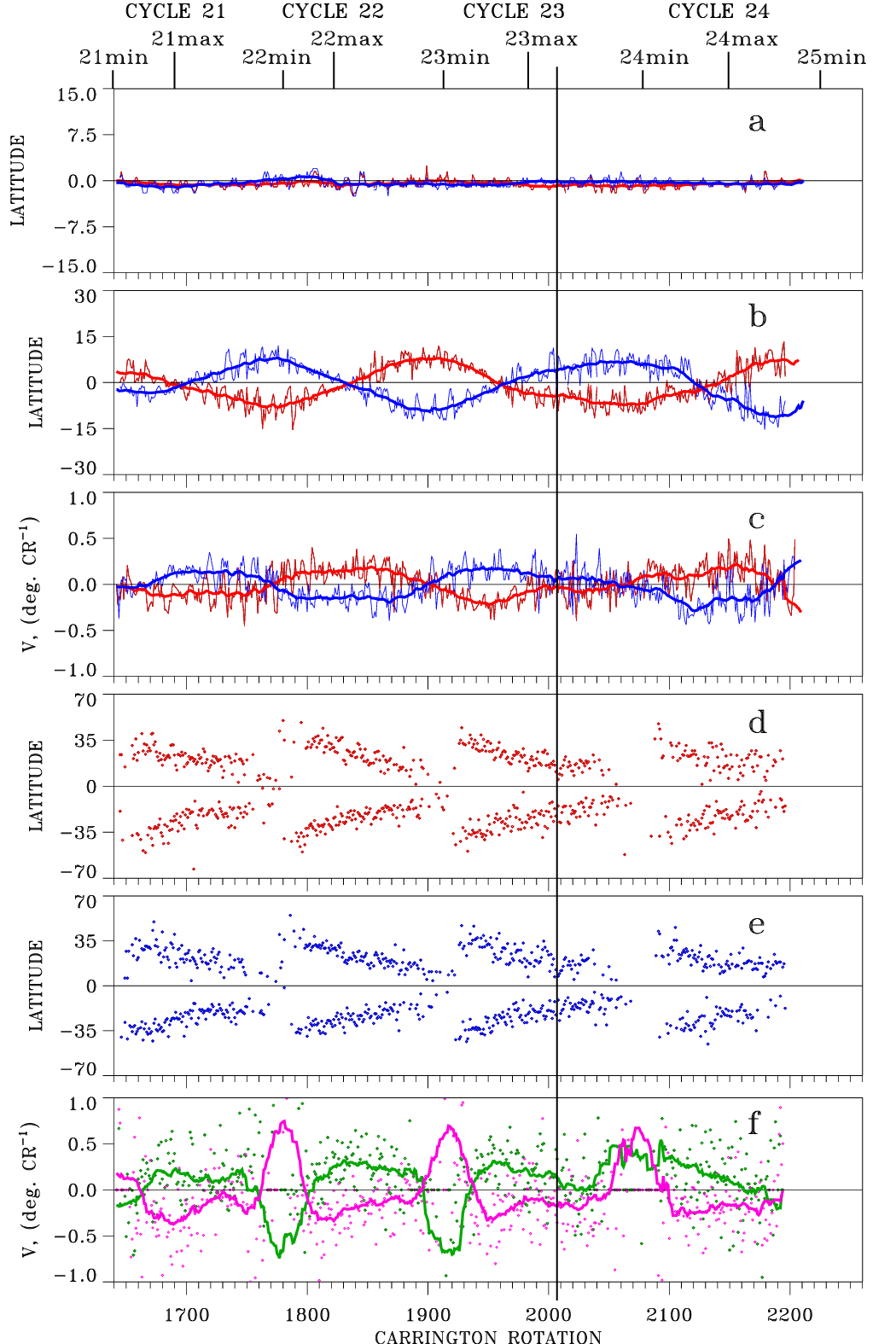}}
\caption{Mean latitudes and velocities of different type
       meridional flows (KPVT-SOLIS).
         Variations in the mean latitudes of low-strength positive- and negative-polarity
         magnetic fields in the range of $0<\mid B \mid\le 7$~G  (a).
   (b)   Variations in the mean latitudes of medium-strength positive-
         and negative-polarity magnetic fields in the range of $ 15 <\mid B \mid\le 25 $~G.
   (c) Velocities of medium-strength positive-polarity (red) and negative-polarity (blue)
         magnetic field flows.
         Variations in the mean latitudes of high-strength positive-polarity (d) and
         negative-polarity (e) magnetic fields in the range of $\mid B \mid>500 $~G.
   (f)    Velocities of the high-strength positive-polarity magnetic field flows
           in the North (green) and  South (lilac) hemispheres.
       Designations are the same as in Figure~\ref{lat_wso}.}
	\label{lat_kp_solis}
\end{figure}

\begin{figure}
\centerline{\includegraphics[width=1\textwidth,clip=]{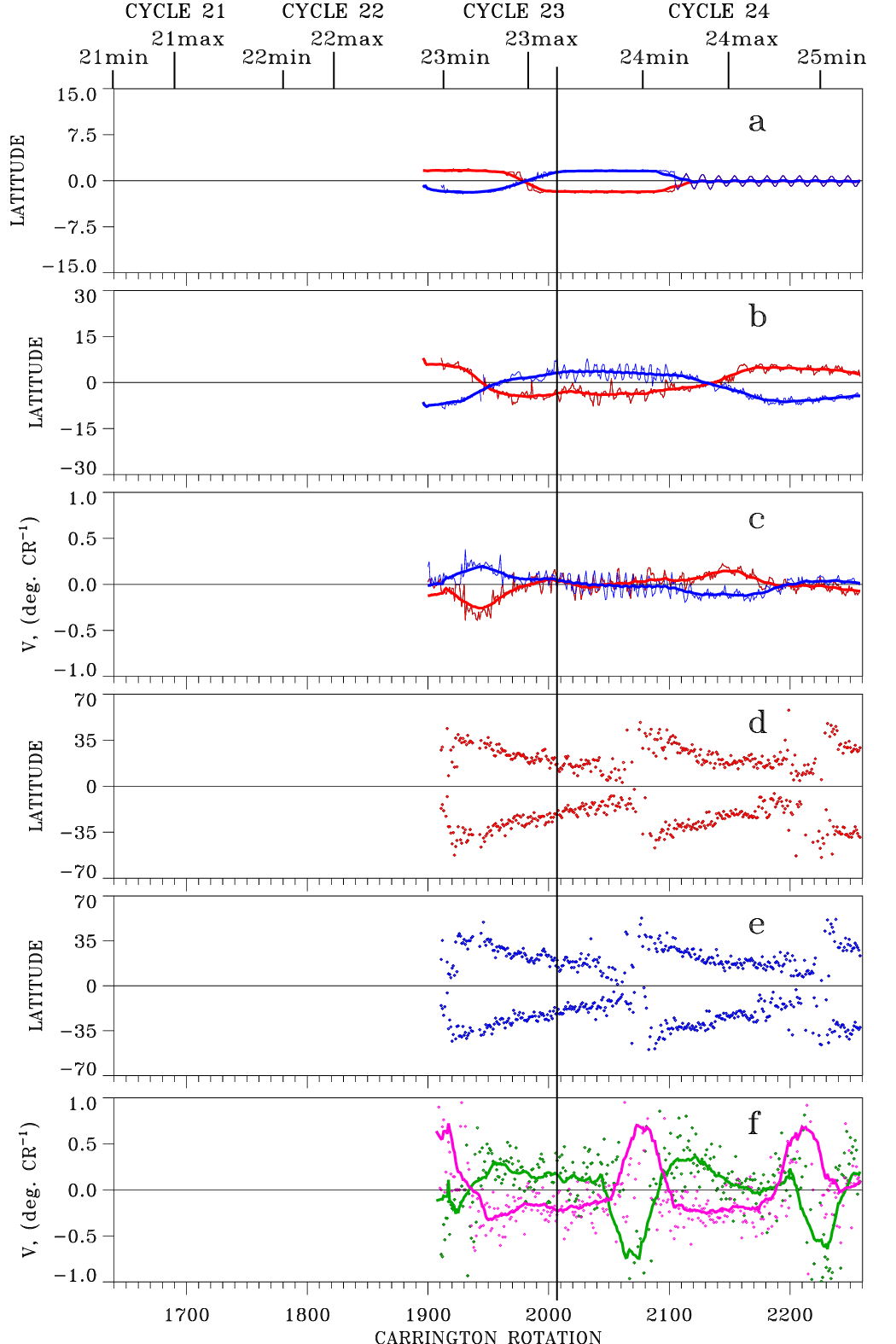}}
\caption{Mean latitudes and velocities of different type
       meridional flows (MDI-HMI).
         Variations in the mean latitudes of low-strength positive- and negative-polarity
         magnetic fields in the range of $0<\mid B \mid\le 7$~G  (a).
   (b)   Variations in the mean latitudes of medium-strength positive-
         and negative-polarity magnetic fields in the range of $ 30 <\mid B \mid\le 50 $~G.
   (c) Velocities of medium-strength positive-polarity (red) and negative-polarity (blue)
         magnetic field flows.
         Variations in the mean latitudes of high-strength positive-polarity (d) and
         negative-polarity (e) magnetic fields in the range of $\mid B \mid>500 $~G.
   (f)   Velocities of the high-strength positive-polarity magnetic field flows
           in the North (green) and  South (lilac) hemispheres.
          Designations are the same as in Figure~\ref{lat_wso}.}
	\label{lat_mdi_hmi}
\end{figure}

Figures~\ref{lat_wso}\,--\,\ref{lat_mdi_hmi}
show the variations in the mean latitudes and velocities of different type
meridional circulations in Solar Cycles 21\,--\,24.
The mean latitudes were calculated for each CR from the corresponding
time-latitude distributions of magnetic fields.
Variations in the mean latitudes of low-strength positive- and negative-polarity
magnetic fields are shown in Figures~\ref{lat_wso}(a), \ref{lat_kp_solis}(a),
and \ref{lat_mdi_hmi}(a).
Their mean latitudes were near the equator.
Positive- and negative-polarity magnetic fields varied in anti phase,
but their variations had no any cycle dependence.

Variations in the mean latitudes of the second type meridional
circulation flows are shown in
Figures~\ref{lat_wso}(b),~\ref{lat_kp_solis}(b),~and~\ref{lat_mdi_hmi}(b).
Wave-like flows were located at low latitudes during solar activity maxima.
The average latitude of the meridional magnetic flows of both polarities
increased with a decrease in solar activity to the minimum of each cycle
in the northern and southern hemispheres.
Waves of positive and negative polarity remained at high latitudes
in each hemisphere until the rising phase of the next cycle.
It should be noted, that waves of the second type meridional flows
occupied a wide range of latitudes in each CR.
Therefore, variations in their mean latitudes indicate
the general direction of the wave migration.

In Figures~\ref{lat_wso}(c),~\ref{lat_kp_solis}(c),~and~\ref{lat_mdi_hmi}(c)
the velocity of the second type meridional flows
in degrees per CR  are shown.
Velocities were higher when meridional the flows were at low latitudes and
they crossed the equator migrating from the North (South)
to the South (North) hemisphere during solar activity maximum.
Velocities decreased to minimal values during the periods when
the flows were at high latitudes during solar activity minimum.
In the polar regions, the meridional flows seem
to turn around and their velocity decreases to zero.

Figures~\ref{lat_wso}(d,~e),~\ref{lat_kp_solis}(d,~e),~and~\ref{lat_mdi_hmi}(d,~e)
show variations in the mean latitudes of the third type meridional circulation,
i.e. the meridional circulation of high-strength positive- and
negative-polarity magnetic fields.
Their meridional flows reflect the well known solar cycle dynamics
of active regions.
They appeared at high latitudes during the rising phases of
solar activity and drifted towards low latitudes.
In Figures~\ref{lat_wso}(f),~\ref{lat_kp_solis}(f),~and~\ref{lat_mdi_hmi}(f)
the  velocities (degree per CR) of the third type meridional
circulation flows are presented.
The latitudes of meridional flows of high-strength positive- and negative-polarity
magnetic fields are almost identical.
When their latitudinal distributions are superimposed, they coincide
with each other with great accuracy.
Therefore the velocities of positive-polarity magnetic fields
in the North and South hemispheres are presented.
The maximal velocities were at the rising solar cycle phases
in all cycles. With cycle  activity the velocities decreased.
Comparison of meridional circulation of medium-strength magnetic fields
with that of high-strength  shows that active regions
did not form when the second type meridional flows were
at the highest latitudes in each cycle.
Active regions began to form when the latitudes of antiphase
meridional flow waves of medium-strength magnetic fields
shifted to lower latitudes as they drifted to the opposite hemispheres.
The formation of active regions stopped when the wave meridional
flows of medium-strength positive- and negative-polarity magnetic
fields migrated away from the equator and approached the opposite poles.

The results indicate that both low-resolution large-scale  (WSO) and
high-resolution  (KPVT-SOLIS and MDI-HMI)
magnetic field data show similar solar cycle changes in the
mean latitude variations of different strength magnetic fields
and the velocities of their meridional circulations.

It should be noted, that the WSO mean latitudes were shifted
to the South hemisphere  in all ranges. It seems to be the instrumental effect.
There were no such shift neither in KPVT-SOLIS  nor in MDI-HMI data.

\section{Discussion}
   \label{S-discussion}

Three different types of  time-latitude distributions and
meridional circulations of solar magnetic fields were revealed.
They depend on the strength of the photospheric magnetic fields.
Both low-resolution large-scale magnetic field data (WSO) and
high-resolution data (KPVT-SOLIS and MDI-HMI)
show the  same three types of photospheric magnetic field dynamics,
but the magnetic field strength values are different due to
different instruments.
The first type includes low-strength magnetic fields.
Low-strength magnetic fields are distributed throughout
the solar surface, but their properties and spatial-temporal variations
are not yet well known.
Currently, it is unclear what mechanism underlies
the origin and dynamics of weak magnetic fields.
It is  unclear whether they are the remnants
of the active region magnetic fields or
the result of a turbulent small-scale dynamo.
In the latter case, their distribution and variations
in the strength of their magnetic field should not coincide
with that in the magnetic fields of active regions.
The small-scale dynamo does not depend on the large-scale dynamo.
The results show that they were distributed evenly across
the solar disk, their time-latitude distribution
and the mean value of their magnetic field strength did not change
during solar cycles from the minimum to  maximum of solar activity.
Their behavior is almost cycle-independent.
This is consistent with the results obtained from daily magnetograms
by \citep{Kleint2010, Buehler2013, Lites2014, Jin2015a, Jin2015b}.
Thus, low-strength magnetic fields show  little variation over
space and time that was not coincide with that of active regions,
indicating that they are predominantly governed
by a process that is independent of the active region solar cycle.
The results indicate that low-strength magnetic fields are not
a product of the decay of active region magnetic fields.
This supports the notion that low-strength magnetic fields
caused by a turbulent small-scale dynamo.

The second type meridional circulation is demonstrated
by medium-strength magnetic fields.
Positive- and negative-polarity magnetic fields reveal
wavy antiphase meridional flows from pole to pole with a period
of approximately 22 years.
They cross the equator during the maximum of solar activity.
Cycle evolution of coronal holes (CHs) revealed
some temporal and spatial regularities
\citep{Bilenko2002, Bilenko2016} that match well the
wave-like pole-to-pole meridional circulation
of medium-strength magnetic fields.
In \cite{Bilenko2002} the pole-to-pole antiphase meridional drift
was revealed for CHs associated with
positive- and negative-polarity photospheric magnetic fields.
In \cite{Bilenko2016} we found two different type
waves of non-polar CHs.
The first are short waves that trace the poleward
movement of unipolar magnetic fields from approximately
$35^{\circ}$ latitude to the corresponding pole in each hemisphere.
The second type of non-polar CH waves forms two
antiphase sinusoidal branches
associated with the positive- and negative-polarity photospheric magnetic fields
with a period of $\approx$268 CRs (22 years).
These CH waves completely coincide with the meridional flows
of the medium-strength positive- and negative-polarity photospheric magnetic fields
revealed in Sections~\ref{S-butterfly_WSO}~and~\ref{S-butterfly_high_res}.
The study of CHs by other authors confirmed our results.
\citet{Pevtsov2010} described the observational evidence
of a transport of one CH across the
equator from one solar polar region to the other.
\cite{Huang2017} found that during the rising phase,
the opposite directed open magnetic fluxes (CH regions),
showed pole-to-pole trans-equatorial migrations in opposite directions.
\cite{Maghradze2022} also found sinusoidal cyclic migration
of the centers of CH activity with opposite directed magnetic
fluxes from pole to pole across the solar equator.

CHs are considered to be good traces of the solar
large-scale, global magnetic field cycle evolution
\citep{Stix1977, Bagashvili2017, Bilenko2017, Bilenko2020}.
Thus, medium-strength magnetic fields reflect
the solar global magnetic field cycle evolution.
Cycle changes in the latitude location  of the CH waves
coincide with that of the axisymmetric component of the solar
dipole ($g_1^0$) \citep{Bilenko2016}.
When both CH branches were above $\approx\pm35^{\circ}$ latitude,
the zonal structure was observed.
When two branches of positive- and negative-polarity
magnetic fields and CHs move below $\approx\pm35^{\circ}$ latitude,
the zonal structure of the GMF changes to sectorial one \citep{Bilenko2016}.
This moment coincides with the beginning of the formation
of high-strength active region magnetic fields.
Apparently, the formation of high-strength magnetic fields
is somehow associated with the second type wave-like
meridional circulation, i.e. medium-strength magnetic field
dynamics.

It is important to note, that \cite{Komm2022} determined
the direction and amplitude of cross-equatorial flows
below the solar surface.
They found that the cross-equatorial flow was mainly southward
during Solar Cycle 23 and mainly northward during Solar Cycle 24.
At depths less than 7 Mm, the average velocity was  -1.1 $\pm$ 0.2 m s$^{-1}$ in
Solar Cycle 23 and +1.3$\pm$0.1 m s$^{-1}$  in Solar Cycle 24.
At the beginning of Solar Cycle 25, the cross-equatorial flow changed sign again.
They concluded, that the subsurface cross-equatorial flow was
nonzero and caused by the inflows associated with active regions
located close to the equator and it was directed to the
hemisphere with the greater amount of flux.
But, apparently, these subsurface meridional flows were associated with
the second type  meridional circulation of medium-strength magnetic fields.
The second type meridional circulation can be associated
with processes occurring at the base of the convective zone.
\cite{Jones2005}, for example, noted that CHs may be rooted as deep as
the base of the convection zone.

The third type is the meridional circulation of high-strength
magnetic fields, i.e. fields of active regions.
Both positive- and negative-polarity high-strength magnetic fields
migrate from high latitudes towards the equator in each cycle.
Neither the magnetic fields of the leading sunspot polarity
nor the following ones migrate to the poles.
It is interesting to note that
using the Kodaikanal Observatory archives for 1906\,--\,1987
and the Mt. Wilson Observatory archives for 1917\,--\,1985,
\cite{Sivaraman2010} found that the latitudinal drifts (or
the meridional flows) of spot groups classified into
three categories of area: 0\,--\,5,
5\,--\,10, and $>$10 millionths of the solar hemisphere
were directed equator ward in both the North
and South hemispheres.
The equator ward drift velocity increases from
almost zero at $35^{\circ}$ latitude in both hemispheres
reaches maximum around $20^{\circ}$ latitude and
slows down towards zero near the equator \cite{Sivaraman2010}.
Consequently, both the active regions themselves and their
magnetic fields migrate from latitudes of $35^{\circ}$
to the equator, and not to the poles.
Some of the high-latitude active region magnetic fields
were captured by the second type wave-like flows
in their drift to the appropriate pole and transported  to the poles.
However, the magnetic fields of active regions are not
the main ones in the process of the solar polar field reversal.

So, the observed magnetic-field cycle dynamics is a
superposition of tree different type time-latitude distributions
and meridional circulations.
Therefore, the solar cycle is governed by different types of dynamo.
\cite{Benevolenskaya1998} proposed a new dynamo model
to explain the solar magnetic cycle that consists
of two main periodic components: a low-frequency
component (Hale's 22 yr cycle) and a high-frequency component
(quasi-biennial cycle).
The essence of the model is that two dynamo sources are at different levels.
The first is located near the bottom of the convection zone,
and the other is near the top.
It is possible that the assumption of the joint operation
of different types of dynamos at different levels
in the convective zone and near the surface will make it possible to better
explain the observed distributions of magnetic fields
of different strengths and their solar cycle dynamics.

\section{Conclusions}
   \label{S-conclusions}

Meridional circulation of the solar magnetic fields  were analyzed
using synoptic magnetic field data from ground based WSO,
NSO KPVT, and SOLIS/VSM and space based SOHO/MDI and SDO/HMI
observatories for Solar Cycles 21\,--\,24.
It have been found that cycle variations of low-, middle-,
and hight-strength magnetic fields significantly different.
Depending on the intensity of the photospheric magnetic fields,
three types of time-latitude distributions and meridional
circulations were identified for both low and high spatial resolution data.

The first type includes low-strength magnetic fields
($ 0<\mid B\mid\le0.6 $~G for WSO  and
$ 0<\mid B\mid\le7$~G  for KPVT-SOLIS and MDI-HMI).
They were distributed evenly across latitude and their average strength
weakly depended on cycle variations of the active region magnetic fields.
These fields are believed to be  determined by a small-scale dynamo.

The second is  medium-strength magnetic field  meridional circulation
($ 0.6 <  \mid B\mid\le 1 $~G for WSO,
$ 10<\mid B\mid\le25 $~G  for KPVT-SOLIS, and
$ 30<\mid B\mid\le70 $~G  for MDI-HMI).
Fields of positive and negative polarity revealed a wave-like
antiphase meridional circulation from pole to pole
with a period of approximately 22 years.
The velocity of meridional flows  were slower
at the minima of solar activity, when they were at high  latitudes
in the opposite hemispheres, and maximal at the solar cycle maxima,
when the positive- and negative-polarity waves crossed the equator.
The flows were more pronounced in large-scale magnetic field data
and in magnetic fields calculated at the source surface.
The meridional circulation of these fields reflects the solar global
magnetic field dynamics and determines the solar polar field reversal.

The third type meridional circulation includes high-strength
magnetic fields, the fields of active regions
($ \mid B \mid >7 $~G for WSO,
$ \mid B \mid >100 $~G  for KPVT-SOLIS, and
$ \mid B \mid >200 $~G  for MDI-HMI).
Magnetic fields of positive and negative polarity were distributed
symmetrically in both hemispheres with respect to the equator.
Active region magnetic fields of both leading and following
sunspot polarity  migrate from high to low latitudes
in all magnetic field ranges.
The velocities of these meridional flows
were higher at the rising and maxima phases than at the minima.

The magnetic fields of active regions are not
the main source of magnetic fields that determine
the solar polar field reversal.
The magnetic fields of some high-latitude active regions
picked up by the meridional wave-like flows (meridional
circulation of the second type) of the corresponding polarity during
the  rising solar cycle phases and transported along with the flows to the poles.

The butterfly diagram is just the result of a superposition
of cyclically changing meridional flows of different strength
positive- and negative-polarity magnetic fields.
The various details and structures in the butterfly diagram,
for example poleward surges,
are the result of the domination of one of the polarities.


\begin{acks}
\noindent Wilcox Solar Observatory data used in this study was obtained via the web site
\textrm{http://wso.stanford.edu} at 2023:04:28\_12:04:38 PDT courtesy of J.T. Hoeksema.
The Wilcox Solar Observatory is currently supported by NASA.

NSO/Kitt Peak magnetic data used here are produced cooperatively
by NSF/NOAO, NASA/GSFC, and NOAA/SEL.

This work utilizes SOLIS data obtained by the NSO Integrated
Synoptic Program (NISP), managed by the National Solar Observatory,
which is operated by the Association of Universities for Research in Astronomy (AURA),
Inc. under a cooperative agreement with the National Science Foundation.

SOHO/MDI data were also used.
SOHO is a project of international cooperation between ESA and NASA.

HMI data are courtesy of the Joint Science Operations Center (JSOC)
Science Data Processing team at Stanford University.	
\end{acks}


%

\bibliographystyle{spr-mp-sola}
\bibliography{bilenko}


\end{article}

\end{document}